\documentclass[preprint2]{aastex}\newcommand{\emul}{false}

\newcommand{\IncludeFigures}{true}

\usepackage{amsmath}
\usepackage{amssymb}
\usepackage{xspace}

\usepackage{ifthen}
\newboolean{emul}%
\setboolean{emul}{\emul}%

\ifthenelse{\boolean{emul}}{
  \usepackage{emulateapj}
  \usepackage{apjfonts}
  \setcounter{secnumdepth}{5}
  \usepackage{graphicx}
  \submitted{Draft \today}
}

\lefthead{FRYER, WOOSLEY, AND HEGER}
\righthead{}
\def\gtaprx {\lower .1ex\hbox{\rlap{\raise .6ex\hbox{\hskip .3ex
	{\ifmmode{\scriptscriptstyle >}\else
		{$\scriptscriptstyle >$}\fi}}}
	\kern -.4ex{\ifmmode{\scriptscriptstyle \sim}\else
		{$\scriptscriptstyle\sim$}\fi}}\xspace}
\def\ltaprx {\lower .1ex\hbox{\rlap{\raise .6ex\hbox{\hskip .3ex
	{\ifmmode{\scriptscriptstyle <}\else
		{$\scriptscriptstyle <$}\fi}}}
	\kern -.4ex{\ifmmode{\scriptscriptstyle \sim}\else
		{$\scriptscriptstyle\sim$}\fi}}\xspace}
\def \sun {\ensuremath{_{\scriptscriptstyle \odot}}\xspace}

\newcommand{\powersep}{{\ensuremath{\times}}}

\newcommand{\g}{{\ensuremath{\mathrm{g}}}\xspace}
\newcommand{\K}{{\ensuremath{\mathrm{K}}}\xspace}
\newcommand{\cm}{{\ensuremath{\mathrm{cm}}}\xspace}
\newcommand{\yr}{{\ensuremath{\mathrm{yr}}}\xspace}
\newcommand{\km}{{\ensuremath{\mathrm{km}}}\xspace}
\newcommand{\Msun}{{\ensuremath{\mathrm{M}_{\odot}}}\xspace}
\newcommand{\Sec}{{\ensuremath{\mathrm{s}}}\xspace}
\newcommand{\erg}{{\ensuremath{\mathrm{erg}}}\xspace}
\newcommand{\ergs}{{\ensuremath{\erg\,\Sec^{-1}}}\xspace}
\newcommand{\ergscc}{{\ensuremath{\erg\,\Sec^{-1}\,\cm^{-3}}}\xspace}
\newcommand{\kms}{{\ensuremath{\km\,\Sec^{-1}}}\xspace}

\newcommand{\gcc}{{\ensuremath{\g\,\cm^{-3}}}\xspace}
\newcommand{\Myr}{{\ensuremath{\mathrm{Myr}}}\xspace}
\newcommand{\Gyr}{{\ensuremath{\mathrm{Gyr}}}\xspace}
\newcommand{\pc}{{\ensuremath{\mathrm{pc}}}\xspace}
\newcommand{\Gpc}{{\ensuremath{\mathrm{Gpc}}}\xspace}
\newcommand{\Mpc}{{\ensuremath{\mathrm{Mpc}}}\xspace}
\newcommand{\kB}{{\ensuremath{k_{\mathrm{B}}}}\xspace}
\newcommand{\ms}{{\ensuremath{\mathrm{ms}}}\xspace}

\newcommand{\Msuns}{{\ensuremath{\Msun\,\Sec^{-1}}}\xspace}

\newcommand{\jeq}{{\ensuremath{j_{\mathrm{eq}}}}\xspace}

\newcommand{\Mbol}{{\ensuremath{M_{\mathrm{bol}}}}\xspace}
\newcommand{\Mdot}{{\ensuremath{\dot{M}}}\xspace}
\newcommand{\Tc}{{\ensuremath{T_{\mathrm{c}}}}\xspace}

\newcommand{\Ye}{{\ensuremath{Y_{\mathrm{e}}}}\xspace}

\newcommand{\vinfall}{{\ensuremath{v_{\mathrm{infall}}}}\xspace}
\newcommand{\vexpand}{{\ensuremath{v_{\mathrm{expand}}}}\xspace}

\newcommand{\lSect}[1]{{\label{sec:#1}}}
\newcommand{\lFig}[1]{{\label{fig:#1}}}

\newcommand{\lTab}[1]{{\label{tab:#1}}}
\newcommand{\pFig}[1]{{\placefigure{fig:#1}}}

\newcommand{\Tabff}[1]{{\ref{tab:#1}}}
\newcommand{\Tab}[1]{{Table~\Tabff{#1}}}

\newcommand{\pan}[1]{{\textit{#1}}}
\newcommand{\Pan}[1]{{Panel~\pan{#1}}}

\newcommand{\FIGFF}[2]{{\ref{fig:#2}\pan{#1}}}

\newcommand{\FIG}[2]{{Fig.~\FIGFF{#1}{#2}}}
\newcommand{\Fig}[1]{{\FIG{}{#1}}}

\newcommand{\Figure}[1]{{Figure~\FIGFF{}{#1}}}

\newcommand{\Sectff}[1]{{\ref{sec:#1}}}
\newcommand{\Sect}[1]{{Sect.~\Sectff{#1}}}
\newcommand{\Sects}[1]{{Sections~\Sectff{#1}}}
\newcommand{\Eqref}[1]{{\ref{eq:#1}}}
\newcommand{\Eqff}[1]{{(\Eqref{#1})}}

\newcommand{\Equation}[1]{{Equation~\Eqff{#1}}}

\newcommand{\Leg}[1]{{\textit{#1}}}

\newcommand{\isofont}[1]{{\mathrm{#1}}}
\newcommand{\isomass}[1]{{\ensuremath{\isofont{^{#1}}}}}
\newcommand{\isocharge}[1]{{\ensuremath{\isofont{_{#1}}}}}
\newcommand{\isotope}[3]{{\ensuremath{\isocharge{#1}\isomass{#2}\isofont{#3}}}}

\newcommand{\I}[2]{{\isotope{}{#1}{#2}}}

\newcommand{\Ep}[1]{{\ensuremath{10^{#1}}}}
\newcommand{\E}[1]{{\ensuremath{\powersep\Ep{#1}}}}

\ifthenelse{\boolean{emul}}{\renewcommand{\pFig}[1]{}}

\newcommand{\FigTwoHundredFiftyFile}{fryfig1.ps}
\newcommand{\FigTwoHundredFifty}{Internal structure of the 250\,\Msun
star at maximum central density as a function of mass coordinate.
\Pan{a} gives the mass fractions of the dominant chemical species.
Note that `iron' denotes the sum of all isotopes of the iron group
elements.  \Pan{b} gives temperature, density, and velocity.  Inward
movement (\vinfall) is drawn as \Leg{dashed} line and outward
movement (\vexpand) as \Leg{dotted} line.  The (logarithmic) density
scale is 3.25 times that of the temperature scale. \lFig{250}}

\newcommand{\FigThreeHundredFile}{fryfig2.ps}
\newcommand{\FigThreeHundred}{Internal structure of the 300\,\Msun
star at a central density of 5.5\E7\,\gcc as a function of mass
coordinate.  \Pan{a} gives the mass fractions of the dominant chemical
species (see \Fig{250}).  \Pan{b} shows temperature, density, and
infall velocity (\vinfall; \Leg{dashed}) at this stage.  \Pan{c} gives
angular velocity ($\omega$) and specific angular momentum in the
equatorial plane (\jeq).  \Leg{Dashed} and \Leg{dash-dotted lines}
give, respectively, the specific angular momentum of the last stable
orbit (LSO) around a Schwarzschild and a Kerr black hole (BH) of that
mass.  The \Leg{dotted line} shows the angular momentum of the LSO
around a BH of mass and total angular momentum enclosed by that mass
shell (see Heger et al.~2000).  \lFig{300}}

\newcommand{\FigDensNRFile}{fryfig3.ps} \newcommand{\FigDensNR}{Density
versus radius of a non-rotating helium core 0, 500, 800 and 930\,\ms
after reaching a central temperature of \Ep{10}\,\K and the transition
to the general relativistic code.  Rapid neutrino cooling in the core
causes the inner 20-25\,\Msun to collapse quickly, decoupling from the
rest of the star.  It is this inner core that first collapses into a
black hole.  Explosive nuclear burning slows the collapse beyond
40\,\Msun, causing the dip in density there.  \lFig{DensNR}}

\newcommand{\FigNeutNRFile}{fryfig4.ps} \newcommand{\FigNeutNR}{Total
energy emitted in neutrinos (\Leg{dotted line}: electron, dashed line:
$\mu$ and $\tau$, \Leg{solid line}: total) as a function of time after
reaching a central temperature of $\Tc=\Ep{10}\,\K$.  As the black
hole grows, the temperature at the last stable orbit decreases,
causing the neutrino emission to decrease dramatically.  Note that
only 1\,\% of the potential energy released in the collapse actually
makes it out of the black hole. \lFig{NeutNR}}

\newcommand{\FigforceFile}{fryfig5.ps} \newcommand{\Figforce}{ Ratio of
centrifugal force over gravitational force at the end of the KEPLER
run (\Leg{dotted line}) and 1.5\,\Sec later, at the end of the 1D
collapse simulation.  The centrifugal force slows the collapse and the
density at the end of the 1D rotating simulation is only
5\E{10}\,\gcc.  Without rotation, the core would have already
collapsed to form a black hole. \lFig{force}}

\newcommand{\FigvelFile}{fryfig6.ps} \newcommand{\Figvel}{Velocity versus
radius 250\,\ms after mapping the simulation into the two-dimensional
code (Model B).  Points correspond to particles at different
radii and latitude.  Thermal pressure, aided by support from
angular momentum, slows the collapse of the core, causing a weak
``bounce''.  The material along the equator is supported by angular
momentum and collapses much slower than the material along the poles.
Hence, at any given radius, there is a range of velocities (the
slowest speeds correspond to the material along the equator).  Note
the accretion shock developing at 2000\,\km.  \lFig{vel}}

\newcommand{\FigpBHFile}{fryfig7.ps} \newcommand{\FigpBH}{The
proto-black hole 0.5\,\Sec before black hole formation.  Color denotes
temperature in \Ep9\,\K and the vectors represent the direction and
magnitude of the particle velocity.  The simulation was actually only
half of this circle and was reflected about the z-axis for display
purposes. It used a total of 25,000 particles though only $\sim
15,000$ are shown in the figure.  At this time, the proto-black hole
has a mass of roughly 78\,\Msun and size of $\sim 1100\,\km$.
\lFig{pBH}}

\newcommand{\FigdensFile}{fryfig8.ps} \newcommand{\Figdens}{Density
versus enclosed mass for the collapsing core 1.75 (\Leg{red; lower
curve}) and 2.6\,\Sec (\Leg{blue; upper curve}) after reaching a
central temperature of $\Tc=\Ep{10}\,\K$.  The points correspond
to particles at different radii and latitudinal angles.  The
``enclosed mass'' is defined by the mass inside a sphere with same
radius as the particle distance from the center.  Because of 
the range in infall velocities, at any given mass coordinate, 
there is a range of densities.  The proto-black hole
mass increases as material accretes through the accretion shock.
However, neutrino cooling causes the core to contract and the
accretion shock actually moves inward.  These two effects cause the
density to increase dramatically.  \lFig{dens}}

\newcommand{\FigneutrinoFile}{fryfig9.ps}
\newcommand{\Figneutrino}{Neutrino Luminosity as a function of time
from Model B.  The $\mu$ and $\tau$ neutrinos (\Leg{dotted line})
dominate the neutrino emission until black hole formation.  Shortly
after the black hole forms, the event horizon grows beyond the $\mu$
and $\tau$ neutrinosphere (at 2.5\,\Sec) and drastically diminishes
the neutrino luminosity.  The electron neutrinos (\Leg{solid line}) do
not decrease significantly until the black hole expands enough to
produce a cool accretion disk.  \lFig{neutrino}}

\newcommand{\FigrotationFile}{fryfig10.ps}
\newcommand{\Figrotation}{$T/|W|$ and instability growth time vs. mass
in the proto-black hole at two different times for Model B: \Leg{solid
line} - at black hole formation, \Leg{dotted line} - 0.5\,\Sec before
black hole formation.  If $T/|W|$ is greater than $\sim0.14$, the
system may develop secular instabilities.  From 5-40\,\Msun, the
growth time of the instabilities is shorter than the actual simulation
time, and this region may have time to clump before collapse, forming
several smaller black holes before merging into a central, spinning
black hole.  \lFig{rotation}}

\newcommand{\FigdiskFile}{fryfig11.ps} \newcommand{\Figdisk}{The
accretion disk 6.5\,\Sec after black hole formation assuming local
angular momentum conservation (Model B).  Velocities are color coded
in units of 1000\,\kms.  At this time, the accretion rate through the
pole is less than 0.1\,\Msuns.  At this time, only 5200 particles
remain in the simulation (4600 are shown in the figure).  The
\Leg{color} represents radial velocity, showing that a large part of
the disk is now stable.  \lFig{disk}}

\newcommand{\FigaccretionFile}{fryfig12.ps}
\newcommand{\Figaccretion}{Total accretion rates for Model A (dotted
line), Model B (\Leg{dashed line}), Model C (\Leg{dot-dashed line}),
and an analytic estimate assuming matter collapses on a free-fall time
(\Leg{solid line}).  The large dip in the accretion rate between 40
and 80\,\Msun in the non-rotating model (Model A) occurs because of
nuclear burning.  Angular momentum transport slows the collapse of
material just above the disk in Model C, causing the accretion rate to
drop slightly quicker in this simulation than in the simulation where
angular momentum is conserved locally (Model B).  Note that the
accretion rate drops precipitously once it is below 10\,\Msuns and
changing the critical accretion rate from 10 to 1\,\Msuns does not
change the black hole mass by more than 10\,\%.  \lFig{accretion}}

\slugcomment{Draft for ApJ, \today} 

\begin{document}

\title{Pair-Instability Supernovae, Gravity Waves, and Gamma-Ray Transients}
\author{C. L. Fryer, S. E. Woosley, and A. Heger}

\vskip 0.2 in
\affil{Department of Astronomy and Astrophysics \\
University of California, Santa Cruz, CA 95064}
\authoremail{woosley@ucolick.org}

\begin{abstract} 

Growing theoretical evidence suggests that the first generation of
stars may have been quite massive ($\sim100-300\,\Msun$).  If they
retain their high mass until death, such stars will, after about
3\,\Myr, make pair-instability supernovae.  Theoretical models for
these explosions have been studied in the literature for about four
decades, but very few of these studies have included the effects of
rotation and none ever employed a realistic model for neutrino
trapping and transport. Both turn out to be very important, especially
for those stars whose cores collapse into black holes (helium cores
above about 140\,\Msun).  We consider the complete evolution of two
zero-metallicity stars of 250 and 300\,\Msun.  Despite their large
stellar masses, we argue that the low-metallicities of these stars
result in negligible mass-loss.  Evolving the stars with no mass-loss
and including angular momentum transport and rotationally induced
mixing, these two stars produce helium cores of 130 and 180\,\Msun
respectively.  Products of central helium burning (e.g. primary
nitrogen) are mixed into the hydrogen envelope, which can dramatically
change the expansion of the envelope, especially in the case of the
300\,\Msun model.  Explosive oxygen and silicon burning cause the
130\,M\sun helium core (250\,\Msun star) to explode, but explosive
burning is unable to drive an explosion in the 180\,\Msun helium core
and it collapses to a black hole.  For this star, the calculated
angular momentum in the presupernova model is sufficient to delay
black hole formation and the star initially forms a 50\,M\sun,
1000\,km core within which neutrinos are trapped.  Although the star
does not become dynamically unstable, the calculated growth time of
secular rotational instabilities is shorter than the black hole
formation time, and such instabilities may develop.  The estimated
gravitational wave energy and wave amplitude would then be $E_{\rm GW}
\approx \Ep{-3}\,\Msun\,c^2$ and $h_+ \approx \Ep{-21}/d(Gpc)$, but
these estimates are very rough and depend sensitively on the
non-linear nature of the instabilities.  After the black hole forms,
accretion continues through a disk.  The mass of the disk depends on
the adopted viscosity, but may be quite large, up to 30\,\Msun when
the black hole mass is 140\,\Msun. The accretion rate through the
disk can be as large as 1-10\,\Msuns.  Although the disk is far too
large and cool to transport energy efficiently to the rotational axis
by neutrino annihilation, it has ample potential energy to produce a
\Ep{54}\,\erg jet driven by magnetic fields. The interaction of this
jet with surrounding circumstellar gas may produce an energetic
gamma-ray transient, but given the redshift and time scale, this is
probably not a model for typical gamma-ray bursts.

\end{abstract}



\section{Introduction}
\lSect{intro}

Simulations of the collapse of primordial molecular clouds suggest
that the first generation of stars (Ostriker \& Gnedin 1996) contained
many extremely massive members, from one hundred to several hundred
solar masses (e.g. Larson 1999; Bromm, Coppi, \& Larson 1999; Abel,
Bryan, \& Norman 2000). While details of the mass function and the
interaction of these stars with their environment have yet to be
worked out, up to 1\,\% of the baryonic mass of the universe might have
participated in this generation of stars (Abel, private communication).  Such
stars (M $\gtaprx$ 100\,\Msun) will reach carbon ignition with helium
core masses in excess of about 45\,\Msun and will encounter the
electron-positron pair instability, igniting carbon and oxygen burning
explosively (e.g., Barkat, Rakavy, \& Sack 1967; Woosley \& Weaver
1982; Bond, Arnett \& Carr 1984; Carr, Bond, \& Arnett 1984; 
Glatzel, El Eid, \& Fricke 1985;
Woosley 1986; Heger and Woosley 2000).  If explosive oxygen
burning provides enough energy, it can reverse the implosion in a giant
nuclear powered explosion. As the mass of the helium core increases, so
does the strength of the explosion and the mass of \I{56}{Ni}
synthesized.  Indeed masses of \I{56}{Ni} ejecta over 40\,\Msun and explosion
energies approaching \Ep{53}\,\erg are possible, with light curves
brighter than \Ep{44}\,\ergs for several months.  Truly these are
``hypernovae'' (Woosley \& Weaver 1982; Paczynski 1998). But they have
not hitherto been associated with gamma-ray transients (GRTs) because
no relativistic matter is ejected.

However, going to still more massive helium cores (over 140\,\Msun) a
new phenomenon occurs as a sufficiently large fraction of the center of
the star becomes so hot that the photodisintegration instability is
encountered before explosive burning reverses the implosion.  This uses up all
the energy released by previous burning stages, and, instead of 
producing an explosion, accelerates the collapse.  A massive
black hole is born inside the star. This has been known for some time,
but here we consider what might occur if, in addition to its high
mass, the helium core is endowed with a moderate amount of
rotation. We follow zero metallicity stars from the main sequence
where they are assumed to rotate rigidly with a ratio of surface
centrifugal force to gravity of 20\,\% - comparable to what is seen in
O-stars today.  By combining the angular momentum transport physics
described in Heger, Langer, \& Woosley (2000) up until stellar
collapse with the two-dimensional rotating core-collapse code
developed by Fryer \& Heger (2000), we are able to follow the entire
life and death of this rotating 300\,\Msun star.  Bond et al. (1984) 
discussed the effects of rotation on the collapse of these massive 
stars and we compare our simulations with their estimates for neutrino 
emission and gravitational wave emission.

Because of their high angular momentum, such massive helium-depleted
cores do not immediately collapse to black holes, but instead form a
hot dense, neutronized core of $\sim$50\,\Msun which continues to
accrete matter for nearly 1\,\Sec before collapsing inside its event
horizon.  Although this ``proto-black hole'' is not dynamically
unstable, it rotates sufficiently rapidly to develop secular
instabilities that may cause it to break into multiple cores, each of
which collapses to a black hole and then merges with the rest.  Even
without this fragmentation, the collapse of a 300\,\Msun star turns
out to be a strong source of gravitational waves with energies and
wave amplitudes of: $E_{\rm GW} \approx \Ep{-3}\,\Msun\,c^2$ and $h_+
\approx \Ep{-21}/d(Gpc)$.

After black hole formation and following the collapse of roughly
another 100\,\Msun, the outer layers of the helium core have
sufficient angular momentum to hang up in a disk which accretes onto
the $\sim 140\,\Msun$ black hole at rates in excess of 1\,\Msuns.
This black hole accretion disk system may produce a new class of
collapsar-like (MacFadyen \& Woosley 1999; MacFadyen, Woosley, \&
Heger 2000) GRTs.  Although neutrino annihilation is too inefficient
to produce polar outflows in such large stars, if magnetic fields can
extract $\sim 1$\,\% of the rest mass of the accreted material and
focus it into $\sim 1$\,\% of the sky, the equivalent isotropic energy
of these GRTs could approach \Ep{56}\,\erg.

\section{Pre-Collapse Evolution of 250 and 300\,\Msun Stars}
\lSect{preSN}

\subsection{Population III Stars}
\lSect{pop3}

It has long been thought that the first generation of stars after the
Big Bang might be characterized by an initial mass function skewed to
more massive stars (e.g., Silk 1983; Carr \& Rees 1984).  More
recently Abel, Bryan, \& Norman (2000) have studied the formation and
fragmentation of primordial, zero metallicity molecular clouds using a
three-dimensional code with adaptive mesh. Dark matter dynamics,
hydrodynamics, and the relevant chemical and radiative processes were
followed down to a scale of 0.5\,\pc.  They concluded that the typical
mass of the first-generation stars is $\sim100\,\Msun$.  Based upon
the amount of energy released by hydrogen fusion, the need to
re-ionize a significant fraction of the universe prior to red shift 5
using the light of such stars, and a crude estimate of the ionization
efficiency of a given ultraviolet photon, one estimates that roughly
0.01\,\% to 1\,\% of the baryonic mass of the universe was
incorporated into such stars. For a \Ep{11}\,\Msun galaxy this
corresponds to $\sim\Ep4$ to \Ep7 massive stars, an estimate that
probably exceeds the rates of merging double neutron star binaries or
black hole binaries in the Milky Way,  $\Ep4-\Ep5$ per
10\,\Gyr (Fryer, Woosley, \& Hartmann 1999).  There is thus ample
theoretical basis for assuming the \emph{formation} of an appreciable
number of primordial stars having mass up to several hundred \Msun.

\subsection{Mass Loss}
\lSect{mdot}

Because of their short lifetime, there are no nearby examples of  
these zero-metallicity, high-mass stars formed in the first generation 
of stars.  Indeed, because the initial mass function of population I and II 
stars is skewed much more toward low-mass stars, there are few examples 
of any nearby massive stars.  Unfortunately, this makes determining 
the mass loss of these stars from observations very difficult.

Mass loss in such heavy stars today is dominated by radiative
processes (Appenzeller 1986) and the stars probably lose most of their
mass before dying.  For zero metallicity, however, as would have
characterized these first stars, radiative mass loss is probably
negligible (Kudritzki 1999).  However, one must still consider mass
loss driven by nuclear pulsations.  The critical mass for the onset of
the nuclear pulsational instability (Schwarzschild \& Harm 1959) is
very uncertain ($90\ldots420\,\Msun$ for stars of solar metallicity;
Appenzeller 1986), and the resulting mass loss rate, even more so.
For very low metallicity, both the temperature dependence of hydrogen
burning and the gravitational potential at the surface of the star
would be different.  Appenzeller's (1970) estimate that a 130\,\Msun
star would lose $\sim30\,\Msun$ during its main sequence lifetime is
probably an overestimate for Pop III stars.  What the value would be
for stars of 200 and 300\,\Msun is unknown, but preliminary
calculations by Baraffe, Heger, \& Woosley (2000) suggest that it
could be small.  These same new calculations also suggest that, at
solar metallicity, opacity and ionization instabilities are more
important in driving mass loss than hitherto realized.

\subsection{Presupernova Evolution of 250 and 300\,\Msun, Z = 0 stars}
\lSect{presn}

We thus consider the evolution of two massive, rotating, Population
III stars of 250 and 300\,\Msun evolved at constant mass.  By
``Population III'', we mean that the initial composition is 76\,\% H
and 24\,\% He with no initial abundance of anything heavier.  The
evolution of these stars is followed using a one-dimensional
hydrodynamic stellar evolution code, \mbox{KEPLER} (Weaver et
al.~1978), from central hydrogen ignition until core collapse.
Stellar rotation is included as described in Heger et al.~(2000).
That is, mixing and redistribution of angular momentum induced by
convection, shear, Eddington-Sweet circulation, etc., are followed, but
the centrifugal force terms are not included in the stellar structure
equations.  Semiconvection is treated as in Woosley \& Weaver (1995),
but no overshoot mixing is assumed.  So long as the ratio of centrifugal
force to gravity remains small, as it did during the epoch followed by
KEPLER, this should be an accurate approximation.  For the initial angular
momentum of the rotating ZAMS stars we assumed rigid rotation with a
velocity 20\,\% Keplerian at the equator, comparable to the typical
value of observed O-stars.  

More details of the evolution of these stars and many other Population III
models will be given elsewhere (Heger \& Woosley 2000).  Here we note
only those characteristics relevant to forming black holes and
producing jets and gravity waves.  After central helium depletion, ($T
= 5\E8\,\K$) the masses of the helium cores in the 250 and 300\,\Msun
models are 130 and 180\,\Msun, respectively.  The hydrogen envelopes
are also appreciably enriched in helium and products of helium burning
with $X = 16.0\,\%$, $Y = 68.0\,\%$, $Z = 16.0\,\%$ for the 250\,\Msun
model and $X = 19.4\,\%$, $Y = 72.2\,\%$, $Z = 8.4\,\%$ (mostly CNO)
for the 300\,\Msun model. The luminosities of the two stars are
3.8\E{40}\,\ergs and 5\E{40}\,\ergs, respectively, and the central
carbon abundance, 6.0\,\% and 4.8\,\%, by mass.

\ifthenelse{\boolean{emul}}{
\vspace{1.5\baselineskip}
\noindent
\includegraphics[width=\columnwidth]{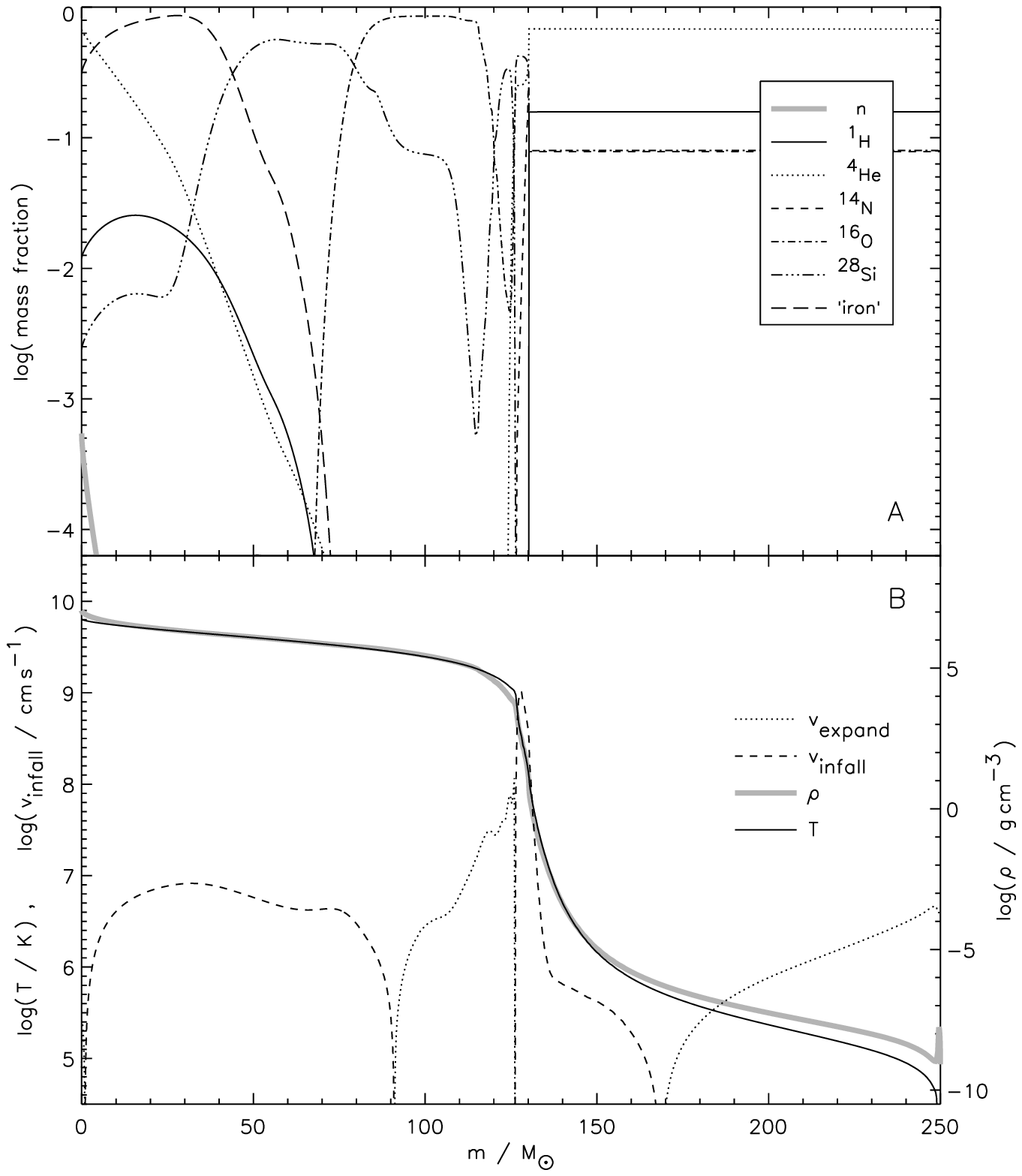}
\figcaption{\FigTwoHundredFifty}
\vspace{1.5\baselineskip}}{}

An interesting characteristic of these two stars is their production
of primary nitrogen.  As a consequence of both the initial
CNO-deficiency and helium-rich nature of their envelopes (resulting
from rotationally induced mixing), the hydrogen burning shell is
characterized by a relatively shallow entropy gradient.  Thus the
envelope remains compact and rapidly rotating with only a small
entropy barrier separating it from the helium core.  During the growth
of the helium-burning convective core, traces of carbon and oxygen are
mixed into the hydrogen-burning shell by meridional circulation.  An
increase of CNO mass fraction to as little as $\gtrsim\Ep{-8}$ is
sufficient to significantly increase the nuclear energy generation
rate and make the hydrogen burning shell convective.  More
significantly, during late stages of helium burning the shear between
the hydrogen shell and the core becomes large enough to lead to a
significant dredge-up of the helium core, i.e., helium and large
amounts of helium burning products - carbon, oxygen, and neon - are
mixed into the envelope.  Because of the short time remaining in the 
life of the star at this point, only part of the carbon and oxygen are 
processed into nitrogen.  For the 250\,\Msun star, this major mixing 
event happens at a central helium mass fraction of $\sim$25\,\% while 
in the 300\,\Msun star it occurs only at the end of central helium burning.  
In the 250\,\Msun star, the mass fraction of primary nitrogen in the 
envelope is 7.75\,\% (i.e., a total of 9.48\,\Msun), while carbon and 
oxygen have mass fractions of 0.26\,\% and 7.97\,\%, respectively.  Since 
in the 300\,\Msun star the dredge-up occurs only towards the end of central
helium burning, the material mixed into the envelope shows a clear
signature of elements produced at the end of hot helium burning.  For
the 300\,\Msun star, the mass fractions are: nitrogen - 1.28\,\% (i.e,
a total of 1.56\,\Msun), carbon - 0.033\,\%, oxygen - 5.68\,\%, neon -
0.964\,\%, and magnesium - 0.395\,\%.

\ifthenelse{\boolean{emul}}{
\vspace{1.5\baselineskip}
\noindent
\includegraphics[width=\columnwidth]{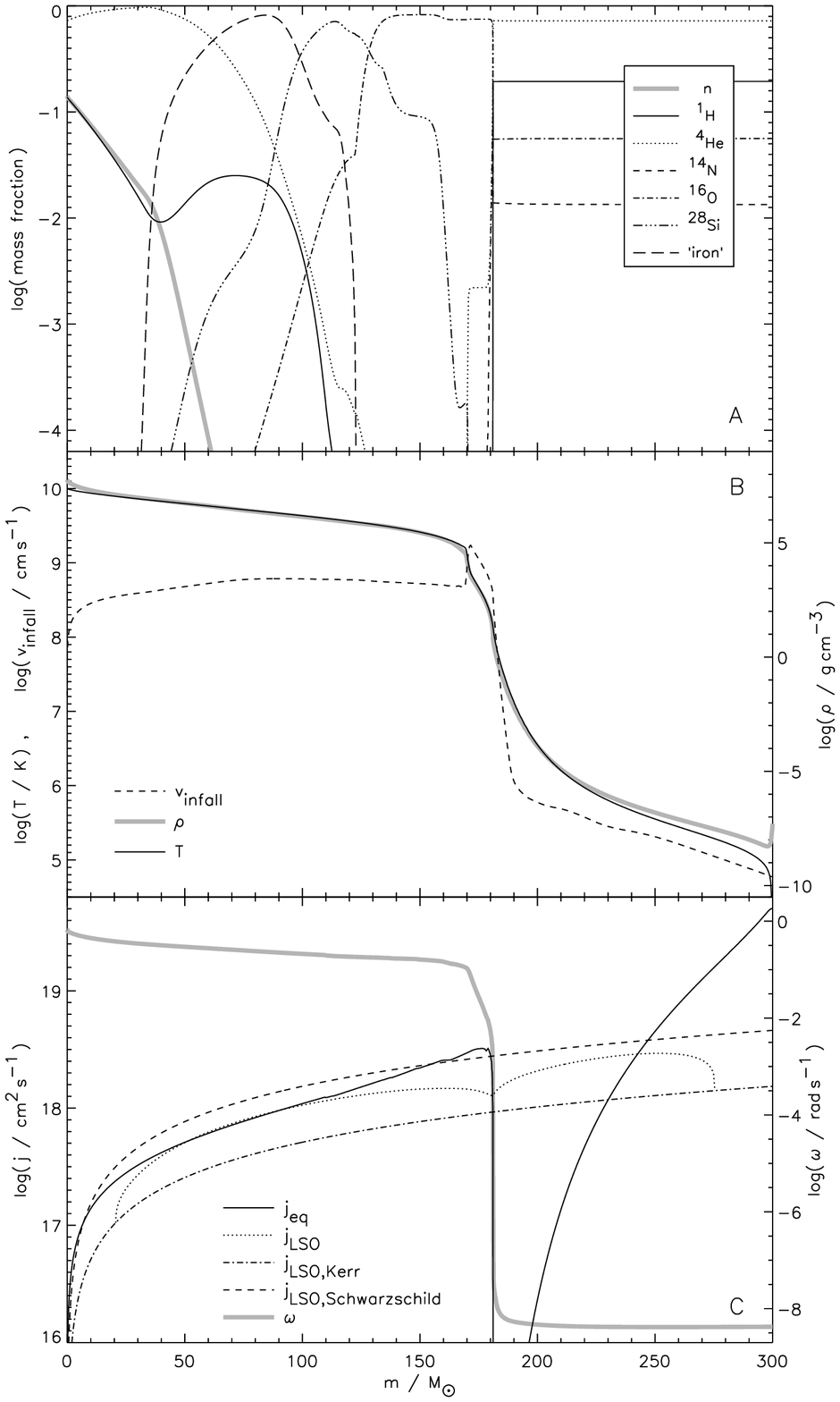}
\figcaption{\FigThreeHundred}
\vspace{1.5\baselineskip}}{}

The increased energy generation afforded by the (suddenly) larger
abundance of CNO leads to the formation of red supergiants in both
stars.  The radius of the 250\,\Msun model expands from 2.3\E{12}\,\cm
to $\sim$\Ep{14}\,\cm by the time it encounters the electron-positron
pair instability; the 300\,\Msun model, from 4\E{12}\,\cm to
$\sim$1.5\E{14}\,\cm.  The mixing events occur $\sim$\Ep5\,\yr before
core collapse in the 250\,\Msun model and $\sim$\Ep4\,\yr before
collapse in the 300\,\Msun model.  Note that all these estimates of
mixing, nitrogen nucleosynthesis, and time-scales are sensitive to the
highly uncertain physics of convection and rotational instabilities.

The new large radius and high metal content in this supergiant phase
can cause significant mass loss and may remove the envelope,
particularly if the star is in a binary.  Once the helium core is
revealed, unless a small amount of hydrogen remains, an even more
pulsationally unstable situation may be created. Appenzeller (1986)
summarizes evidence that bare helium cores above $\sim$16\,\Msun may
be unstable to nuclear driven pulsations.  However, especially for the
300\,\Msun model, the time remaining until the star dies is very
short. If the core can avoid losing an additional 40\,\Msun of helium
(i.e., \Mdot $\ltaprx$ 0.004\,\Msun y$^{-1}$), a black hole will form.

\pFig{250}

\section{Collapse and Black Hole Formation}
\lSect{collapse}

Following helium depletion, both the 250 and 300\,\Msun stars
encounter the electron-positron pair instability. The 250\,\Msun star
experiences a very deep bounce, penetrating into the pair-unstable
region, then violently rebounds because of the excess energy created
by explosive oxygen and silicon burning. The peak temperature and
density during the bounce are 6.37\E9\,\K and 1.08\E7\,\gcc
(\Fig{250}).  So much of the core is heated to temperatures in excess
of $\sim5\E9\,\K$ that 43.0\,\Msun of \I{56}{Ni} is synthesized.  We
estimate that this is very nearly the maximum mass that can explode by
nuclear burning alone and so represents a nearly maximal mass of
$^{56}$Ni and explosion energy, 9\E{52}\,\erg of kinetic energy at
infinity, almost 100 times that of an ordinary supernova.  Heger et
al. (2000) have calculated a peak luminosity for this model of
$\sim$\Ep{44}\,\ergs ($\Mbol=-21$) lasting for about 150 days.
Roughly the same peak luminosity and duration characterizes the event
with and without the hydrogen envelope, but in the case of the helium
core the peak is delayed about 100 days and completely powered by
\I{56}{Co} decay.  Since the kinetic energy of the explosion is so
high, the interaction with circumstellar matter might also be quite
brilliant. These are truly ``hypernovae'' (Woosley \& Weaver 1982),
but no strongly relativistic matter is ejected and this explosion will
not produce a GRT.

\pFig{300}

The 300\,\Msun model (180\,\Msun helium core) on the other hand is so
tightly bound and gains so much kinetic energy prior to oxygen burning
that even the fusion of the entire core to silicon and iron is unable
to reverse its infall.  It must make a massive black hole. At the
latest time reliably calculated with KEPLER, the central temperature
had reached \Ep{10}\,\K and the nuclei in the center of the star had
photodisintegrated to 72\,\% alpha particles and 28\,\% nucleons (\Ye
= 0.50). Unlike what is seen in ordinary supernova models, even though
temperature dependent partition functions were included in the
calculation, the higher entropy of these models results in complete
photodisintegration of heavy elements.  The density at this point is
5.5\E7\,\gcc, the total kinetic energy of infall, 5.8\E{52}\,\erg, and
the net binding energy, $-7.9\E{52}\,\erg$.  \Figure{300} shows the
composition, temperature and density structure, and distribution of
specific angular momentum at this time. In subsequent discussions we
define this to be the t = 0 collapse model.

\subsection{Collapse without Rotation}

Beyond this point the effects of neutrino trapping are no longer
negligible and the Kepler model was mapped into a 
one-dimensional Lagrangian core-collapse code (Herant et al.~1994, 
Fryer et al.~1999) which included the effects of general relativity 
and neutrino transport. It is useful to compare the ensuing collapse of this
300\,\Msun star to the core-collapse of an ordinary 15\,\Msun
supernova progenitor.  Just like ordinary supernovae, as the core of
the 300\,\Msun star contracts, the rate of electron capture increases
and the removal of electron degeneracy pressure and cooling via
neutrino emission help to create a runaway collapse.  However, there
are several important differences between the structures of 300\,\Msun
and 15\,\Msun stars.  As noted previously, the entropy in the more
massive core is larger and thus favors the more complete
photodisintegration of heavy elements and alpha particles.  General
relativity also plays a more significant role.  The collapse of a
15\,\Msun core halts when its central density exceeds a few times
\Ep{14}\,\gcc, i.e., when nuclear forces and neutron degeneracy
pressure become important.  However, the core of a 300\,\Msun star is
so large that it collapses into a black hole before nuclear forces can
affect the collapse, less than 1\,\Sec after the code link at
$\Tc=\Ep{10}\,\K$.

\Figure{DensNR} shows the velocity, electron fraction ($Y_e$),
temperature and density profiles of the 300\,\Msun core at times zero,
500, 800, and 930\,\ms after the code link.  Before the initial
collapse, the entropy of the core is $\sim10\,\kB$ per nucleon
(compare to $\sim 1 - 2\,\kB$ per nucleon for most core-collapse
supernovae).  Hence, for a given central collapse density, the
300\,\Msun star is hotter and neutrino emission is more efficient.
Neutrinos from electron capture stream out from the
core, quickly lowering its lepton number and lowering the electron 
degeneracy pressure.  However, because the core
of a 300 M\sun \ star is so much bigger than a 15 M\sun supernova 
progenitor, electron neutrinos are trapped at a lower central 
density (Bond et al. 1984).  Initially, neutrino absorption dominates 
the electron neutrino opacity, but as the temperature rises, pair 
scattering quickly becomes the most important opacity source.
Although $\mu$ and $\tau$ neutrinos escape the collapsing star 
more easily, trapping the electron neutrinos halts the deleptonization 
in the core.  But at this point, the collapse of the core is already 
inevitable.  The large population of electron neutrinos in the core 
actually causes the central electron fraction to rise, as the core strives 
to reach an equilibrium between the electron/anti-electron neutrinos and the 
electron/proton fractions in the core (see \Fig{DensNR}).

In our simulations, we follow the collapse until a part of the star
falls below its last stable orbit (for the non-rotating star this is
$6 G M_{\rm enclosed}/c^2$).  We use the simplifying assumption that
the core ``forms a black hole'' at this point.  A rigorous
determination of the first trapped surface would require a 
multi-dimensional general relativistic calculation that is 
beyond the scope of this paper.  The exact point in the core where the 
initial black hole forms is also very sensitive to the neutrino transport
(especially upon the $\mu$ and $\tau$ neutrinos which dominate the
neutrino cooling).  In our simulations, the black hole first forms at
$\sim 20\,\Msun$, but the entire inner 35\,\Msun of the core is very
close to its last stable orbit and relatively small changes in the
$\mu$ and $\tau$ transport could give an initial black hole mass
anywhere in the range of 15-35\,\Msun.  However, no matter where the
black hole initially forms, without rotation, most of the helium core
will very quickly become part of the black hole.

The bulk of the potential energy released during the collapse is
dragged into the black hole.  Without rotation to drive a dynamo, it
is unlikely that a strong magnetic field will emerge during the
collapse and magnetic fields will not be able to tap any of the
potential energy.  Likewise, a non-rotating collapse will not emit
much energy in gravitational waves.  Bond et al. (1984) argued that 
much of the potential energy could be released in a neutrino fireball.  
However, their analytic calculation overestimated the temperature at 
the neutrinosphere, which drastically overestimated the total neutrino 
luminosity.  In our simulations, most of the neutrinos are trapped 
in the flow and only 1\,\% of the gravitational potential energy is 
released in neutrinos, most in $\mu$ and $\tau$
neutrinos (\Fig{NeutNR}).  The collapse of a non-rotating 300\,\Msun
star proceeds without so much as a whimper.

\pFig{DensNR}
\ifthenelse{\boolean{emul}}{
\vspace{1.5\baselineskip}
\noindent
\includegraphics[angle=0,width=\columnwidth]{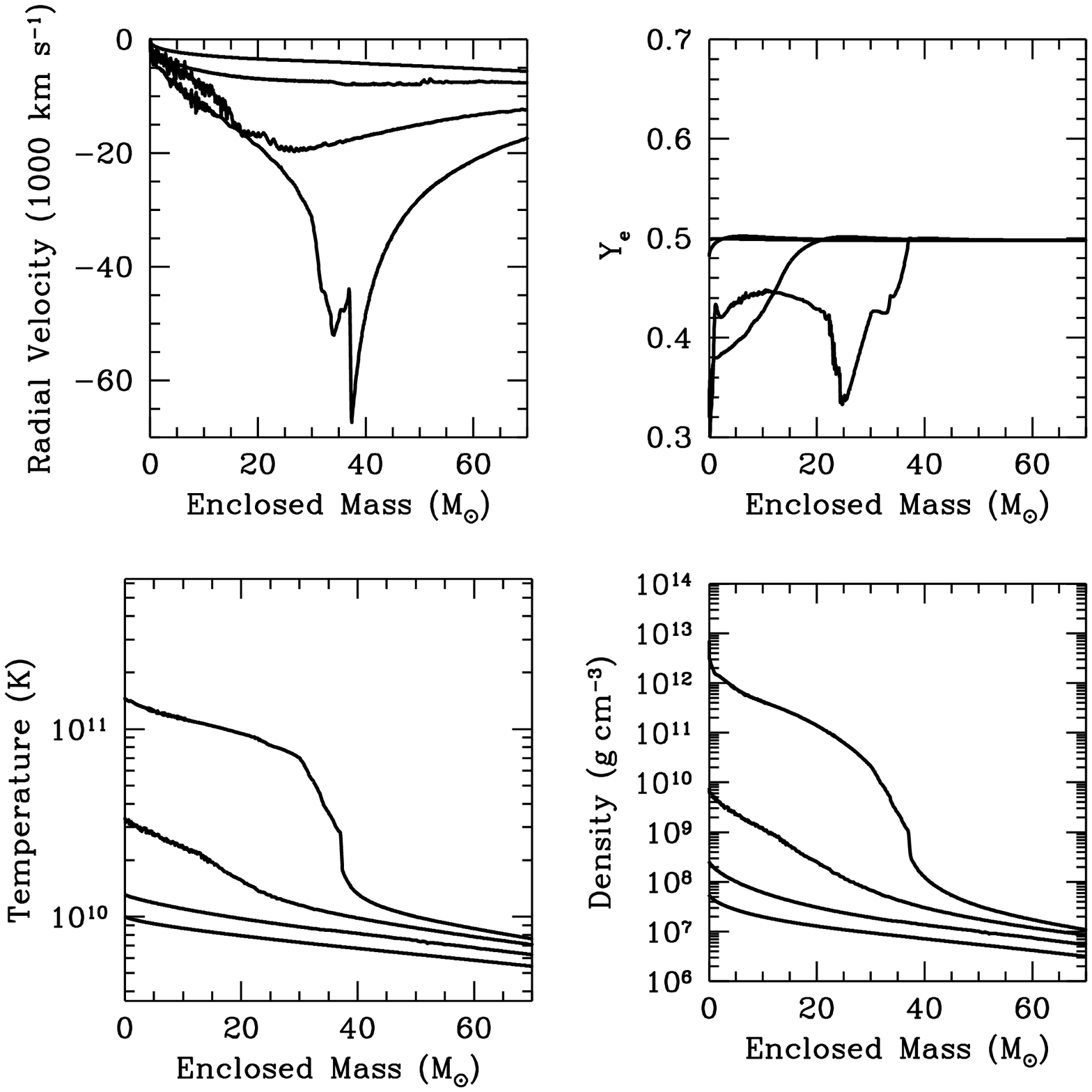}
\figcaption{\FigDensNR}
\vspace{1.5\baselineskip}}{}

\pFig{NeutNR}
\ifthenelse{\boolean{emul}}{
\vspace{1.5\baselineskip}
\noindent
\includegraphics[angle=0,width=\columnwidth]{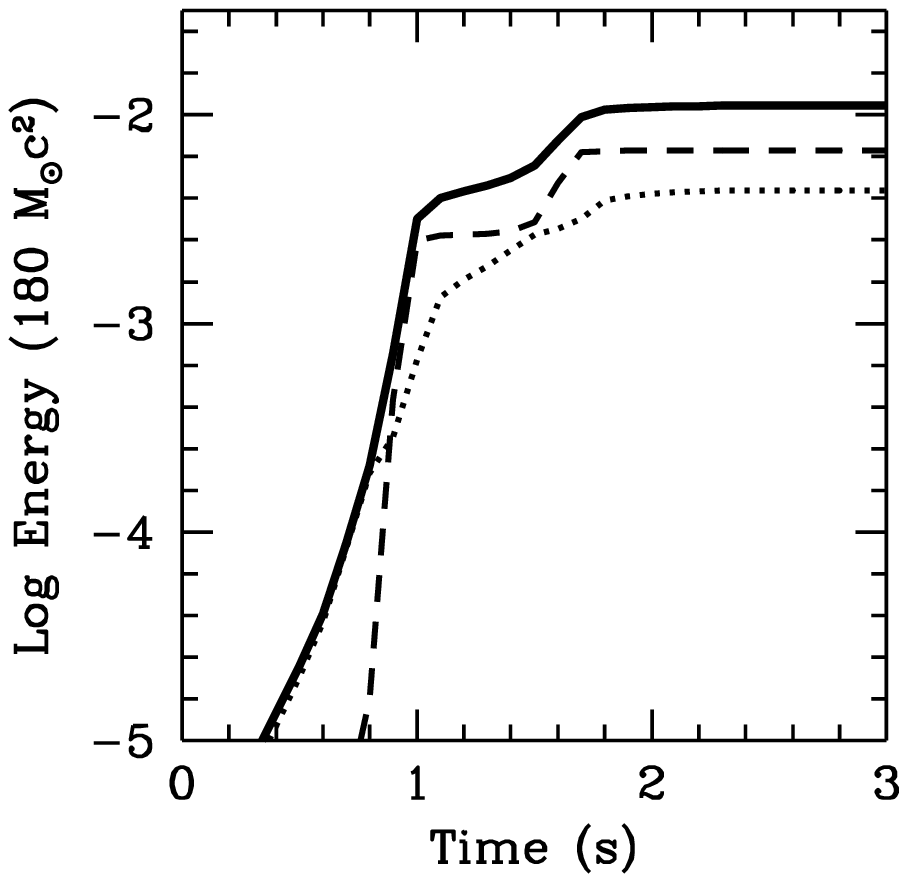}
\figcaption{\FigNeutNR}
\vspace{1.5\baselineskip}}{}

\subsection{Collapse With Rotation}

However, these stars are rotating and this plays an important role in
the collapse.  As with the non-rotating model, the rotating collapse
is simulated by first mapping the KEPLER output into the
one-dimensional neutrino transport code.  The effect of rotation is
included by adding a centrifugal term, $a_{\rm cent}=j^2/r^3$, to the
force equation and conserving angular momentum locally, $j\equiv
j(m,t)=j(m)$, for each mass $m$.  We map $j$ from the KEPLER output
(\Fig{300}) by setting $j(r)=r \, \Omega(r)$ and add the equatorial
force to the entire star.  This approximation obviously overestimates
the effect of angular momentum everywhere except in the equatorial
plane, but allows us to follow the first 1.5\,\Sec of the star's
collapse quickly and gives approximately the correct composition and mass
distribution.  Recall that in the non-rotating simulation, a black
hole formed 1\,\Sec after mapping from the KEPLER output into our 1D
neutrino transport code.  In the rotating simulation, centrifugal
forces become extremely important (\Fig{force}) and slow the collapse.
After 1.5\,\Sec, the central density of the rotating core is only
$5\E{10}\,\gcc$.

\pFig{force}
\ifthenelse{\boolean{emul}}{
\vspace{1.5\baselineskip}
\noindent
\includegraphics[angle=0,width=\columnwidth]{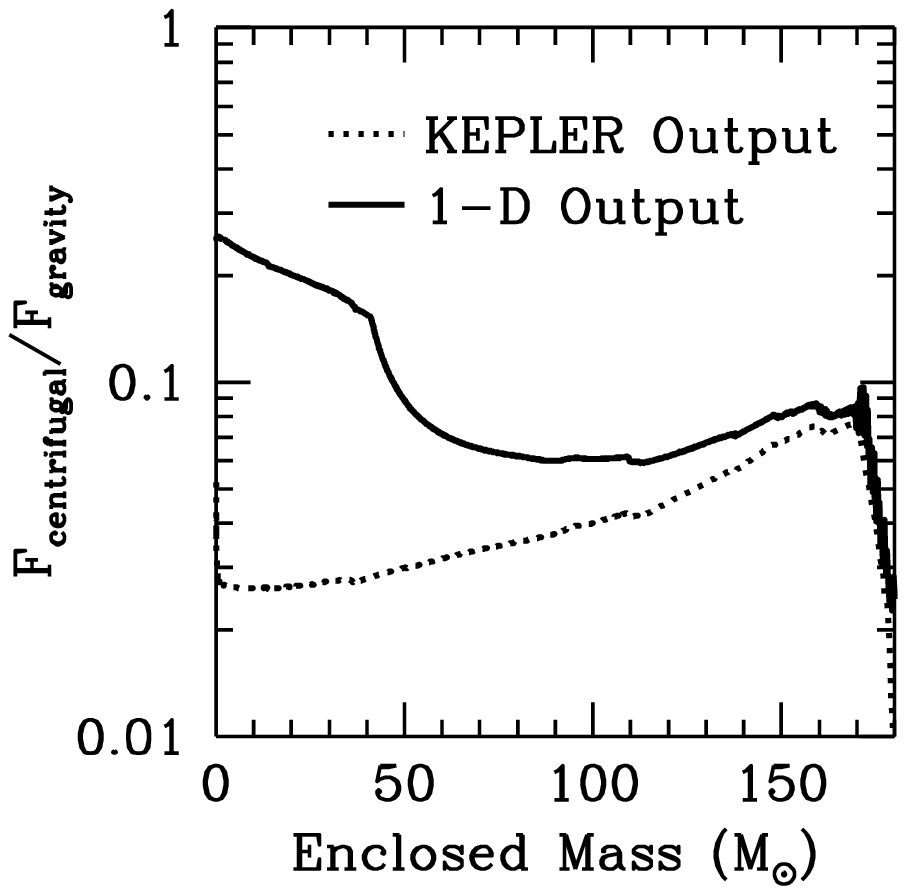}
\figcaption{\Figforce}
\vspace{1.5\baselineskip}}{}

When the centrifugal force exceeded 25\% of gravity for any lagrangian
zone (Fig. 5), the one-dimensional collapse calculation was halted
again and remapped into a two-dimensional code based upon the smooth
particle hydrodynamics prescription (Herant et al.~1994, Fryer \&
Heger 2000). Use of the 2D code was delayed until this point in order
to make the problem computationally tractable.  The rotational axis
was aligned with the code's axis of symmetry and the angular momentum
of each particle was given by the rotation rate of the zones in the
one-dimensional calculation: $j(\mathbf{r})=\Omega(r) \, r \,
\cos(\theta)$ where $\theta$ is the angle above the plane of rotation.
The net effect is to give (initially) spherical shells constant
angular velocity and a total angular momentum equal to what they had
in the Kepler model at t = 0.

A total of three two-dimensional simulations were performed this way,
each employing a different treatment of the angular momentum.  The
models will be referred to by three names: A) no rotation (this is
inconsistent with the assumed density structure, but calculated just
for comparison); B) the star rotates with the angular momentum
distribution at the end of the 1D calculation - which is the same as
prescribed by the KEPLER calculation; in the continued evolution the
angular momentum of each particle is conserved; and C) same as B, but
with angular momentum transport between particles mediated via an
alpha-disk prescription ($\alpha_{\rm disk}\ =0.1$: see Fryer \& Heger
2000 for details).  In the remainder of this section, we discuss the
collapse of the star and the resultant black hole formation under
these three assumptions.  A summary of results is given in
\Tab{300}.

As might be expected from our non-rotating collapse simulation, Model
A collapses directly into a black hole 250\,\ms after the beginning of
the two-dimensional simulation.  As before, we assume that the star
has made a black hole when matter at any radius lies within its last
stable orbit for the given distribution of matter.  At that point, all
particles inside this radius are removed and replaced by a perfectly
absorbing inner boundary condition at that radius.  This way, we can
follow the continued accretion into the black hole (\Sect{accretion}).

\pFig{vel}
\ifthenelse{\boolean{emul}}{
\vspace{1.5\baselineskip}
\noindent
\includegraphics[angle=0,width=\columnwidth]{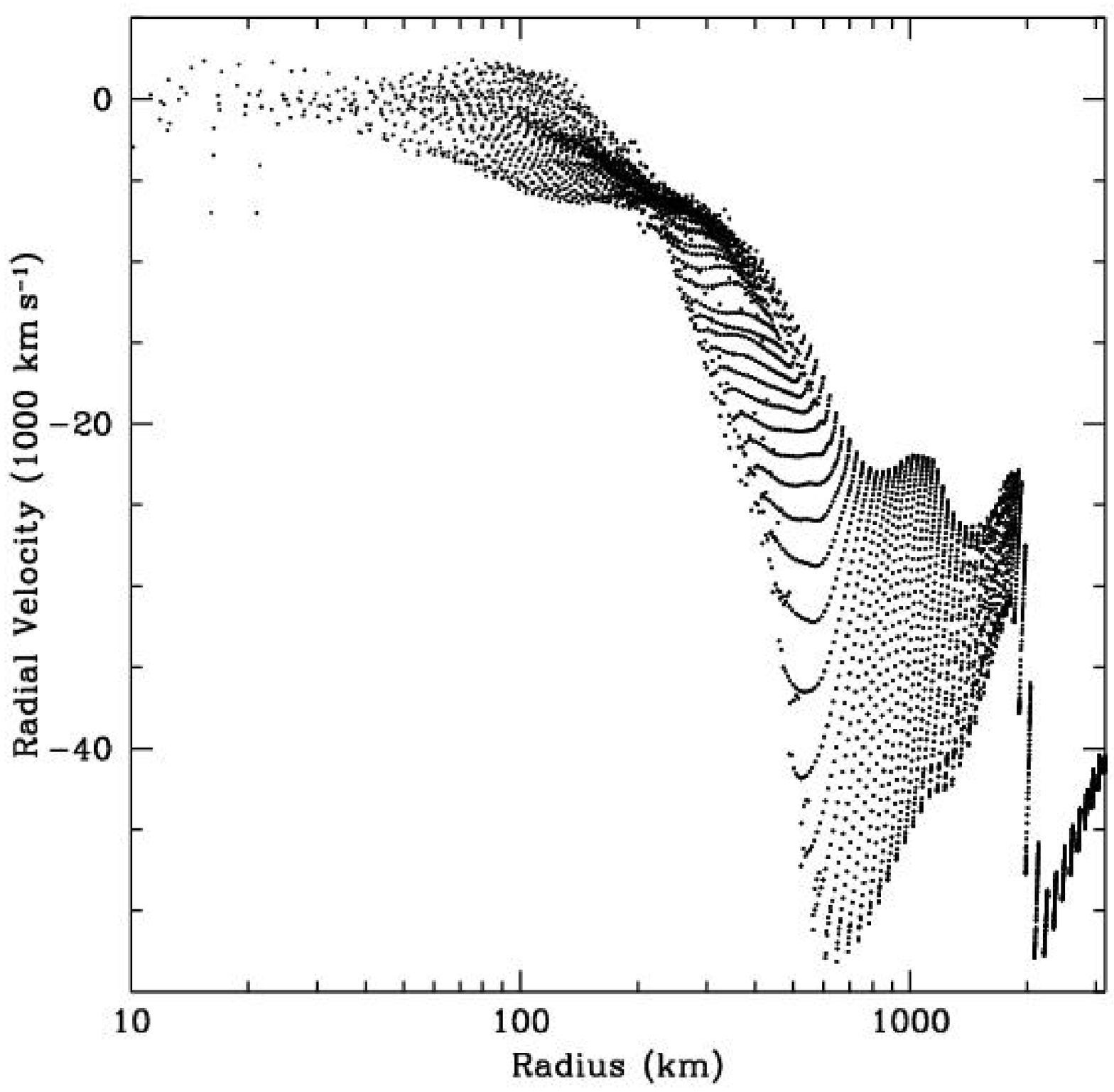}
\figcaption{\Figvel}
\vspace{1.5\baselineskip}}{}

\pFig{pBH}
\ifthenelse{\boolean{emul}}{
\vspace{1.5\baselineskip}
\noindent
\includegraphics[angle=90,width=\columnwidth]{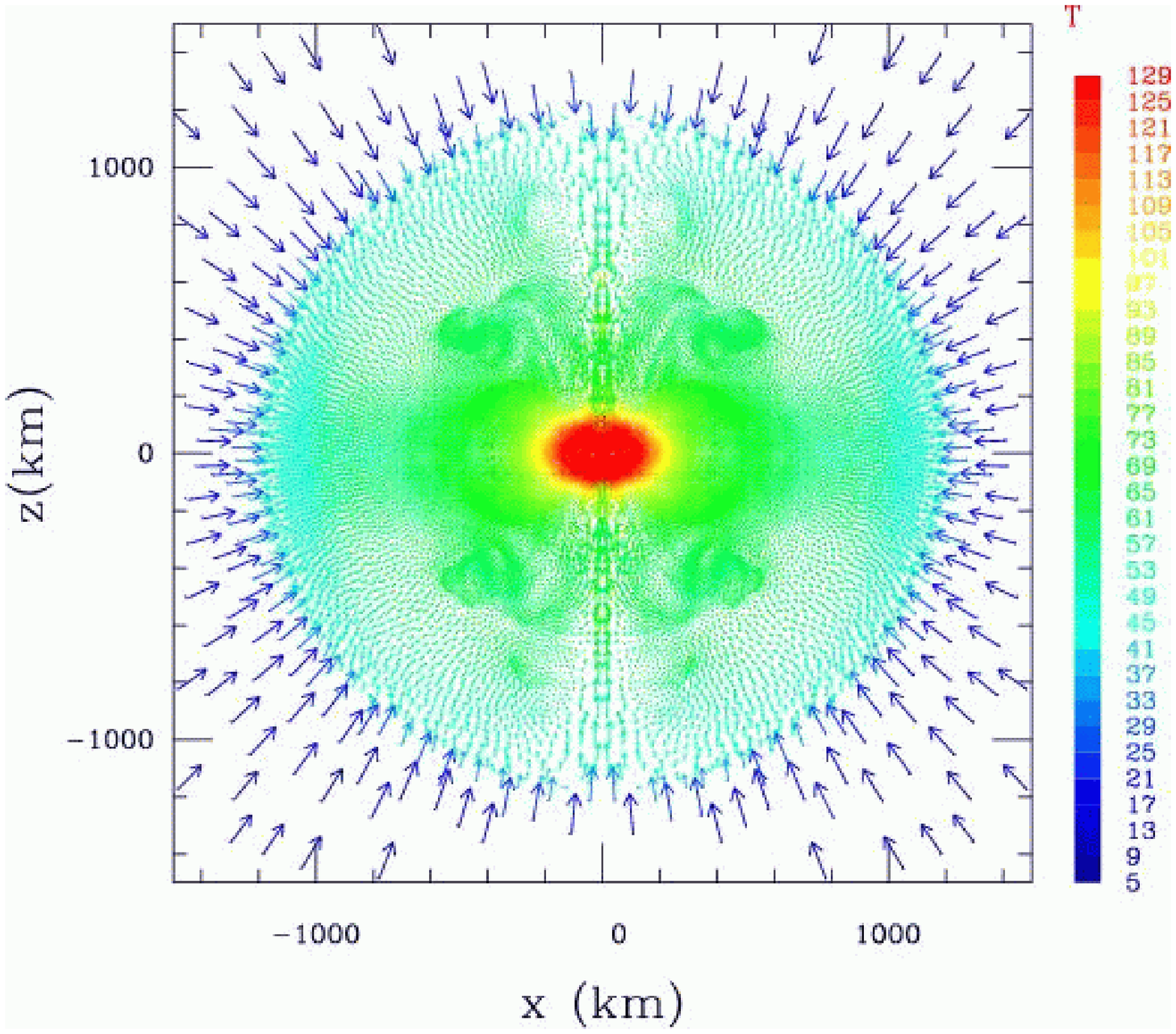}
\figcaption{\FigpBH}
\vspace{1.5\baselineskip}}{}

\pFig{dens}
\ifthenelse{\boolean{emul}}{
\vspace{1.5\baselineskip}
\noindent
\includegraphics[angle=0,width=\columnwidth]{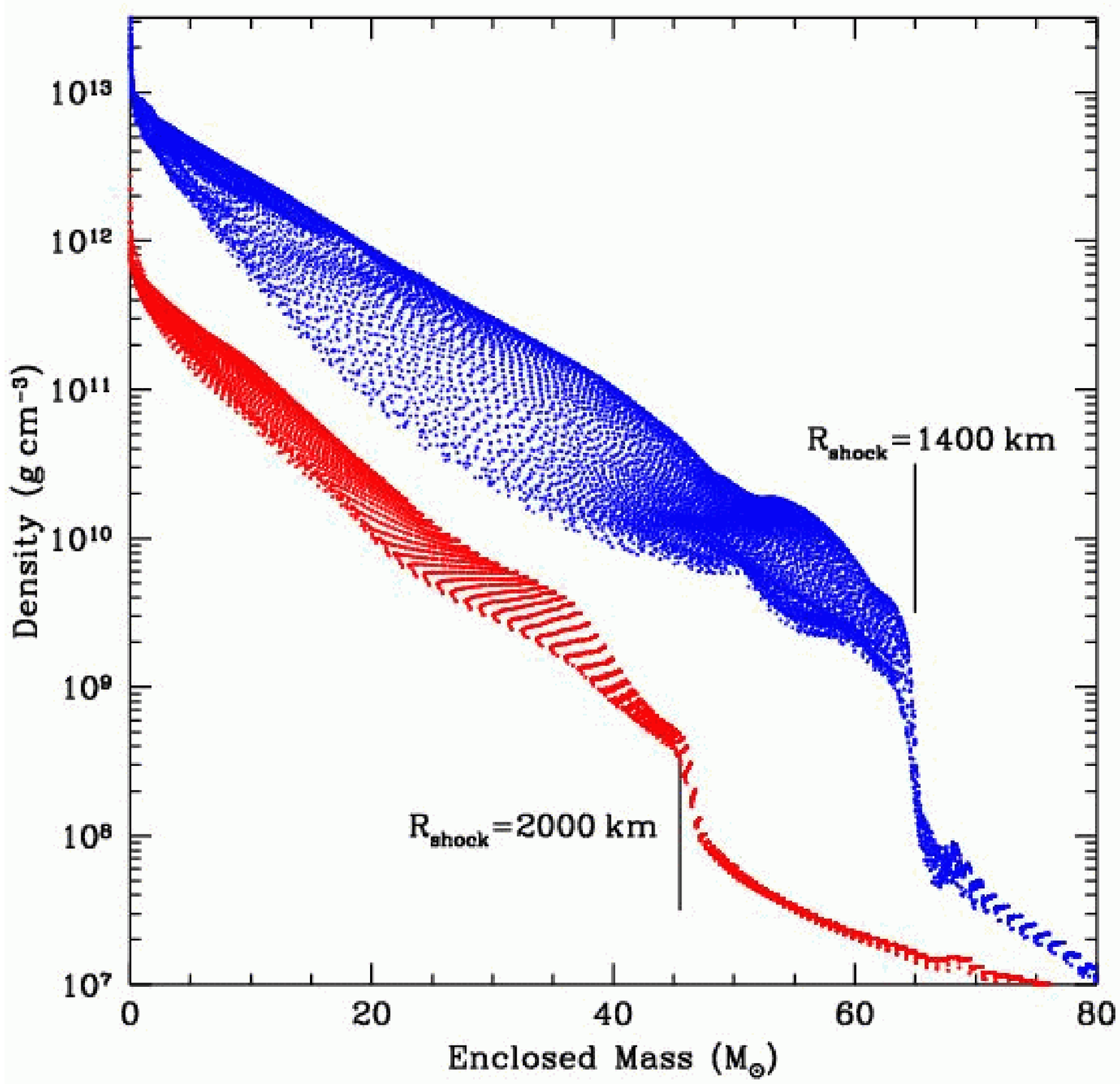}
\figcaption{\Figdens}
\vspace{1.5\baselineskip}}{}

The models which include the KEPLER angular momentum distribution
(\Fig{300}) evolve very differently from the non-rotating case.  As
the star collapses, neutrinos become trapped in the flow, maintaining
$\Ye \sim 0.35$.  The high entropy of the 300\,\Msun star, along with
the rapid rotation, halts the collapse of the core and even produces a
weak ``bounce'' at a central density of only a few times
\Ep{12}\,\gcc (\Fig{vel}).  The structure of the resulting core
is similar to the proto-neutron star cores encountered in ordinary
core-collapse supernovae, but here it is much larger.  Usually, in
lower mass supernovae, after nuclear forces and neutron degeneracy
pressure halt the collapse, the core bounces, sending a shock into the
star.  When the bounce shock stalls, it forms a proto-neutron star
core capped by the accretion shock produced as the remainder of the
star falls onto the core.  In the case of core-collapse supernovae,
the proto-neutron star has a mass of roughly 1\,\Msun and a radius of
$\sim 400\,\km$.  In the present simulations of the 300\,\Msun star,
our weak ``bounce'' yields a 50\,\Msun ``proto-black hole'' with a
radius of 1000\,\km (\Fig{pBH}). By proto-black hole, we mean a hot
dense neutronized core that has not contracted inside its event
horizon, but which would do so if lost its internal entropy.

\pFig{neutrino}
\ifthenelse{\boolean{emul}}{
\vspace{1.5\baselineskip}
\noindent
\includegraphics[width=\columnwidth]{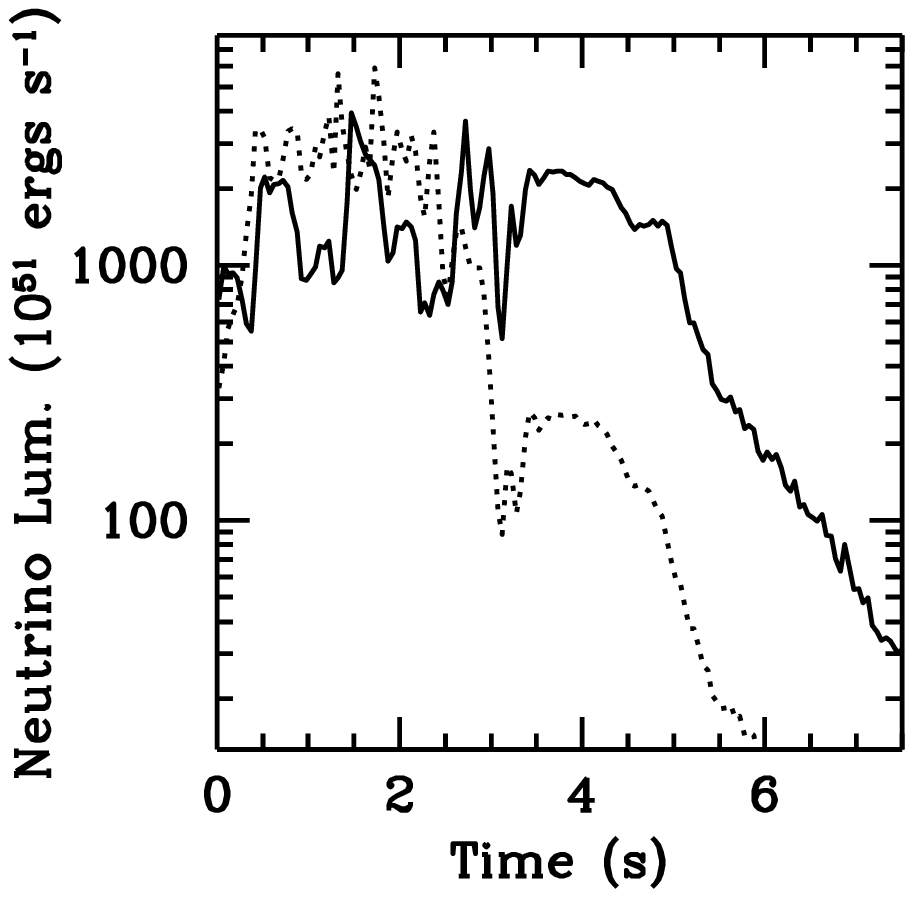}
\figcaption{\Figneutrino}
\vspace{1.5\baselineskip}}{}

Material continues to accrete onto this core through a shock at a rate
of 10-100\,\Msuns.  By 1.1\,\Sec after the beginning of the
two-dimensional simulations (2.6\,\Sec after the initial collapse in
\Fig{300}), the proto-black hole mass has increased to nearly
70\,\Msun (\Fig{dens}).  Neutrino emission, dominated by $\mu$ and
$\tau$ neutrinos (since the electron neutrinos are trapped), cools the
proto-black hole and the radius of the accretion shock slowly shrinks.
From \Fig{neutrino}, we see that the total neutrino
luminosity is well above \Ep{54}\,\ergs during much of the simulation.
Such high neutrino luminosities suggest the possibility of a
supernova-like explosion or perhaps a GRB (Fuller \& Shi, 1999).

However, inside the proto-black hole, scattering and absorption
dominate the neutrino opacity and beyond the $\sim$1000\,\km
neutrinosphere, the density of neutrinos is too low to allow
significant numbers to annihilate.  The neutrino density is roughly
\Ep{30}\,\ergscc, over an order of magnitude lower than the
\emph{lowest} neutrino density considered by Popham, Woosley, \& Fryer
(1999) for black hole accretion disks.  In addition, the bulk of the
neutrinos emitted by the 300\,\Msun model are a) traveling radially
and the neutrino cross-section is only large for head-on collisions
and b) $\mu$ and $\tau$ neutrinos which have lower neutrino
annihilation cross-sections.  Combining all of these effects, neutrino
annihilation injects less than \Ep{47}\,\ergs into the infalling
material.  Given that the binding energy of the material piling up at
the accretion shock is over \Ep{52}\,\erg, we conclude that neutrino
annihilation does not drive an explosion.  Neutrinos interacting with
nucleons and scattering on electrons were also included in the
calculation and these too provided totally insufficient energy to
reverse the infall.

However, shortly after the black hole forms, the accretion shock moves
inwards and angular momentum becomes more important.  An accretion
disk will eventually form and that may produce an explosion.  Thus, it
is important to determine when the black hole forms.  As the
proto-black hole contracts, its inner core is slowly compressed down
to the event horizon.  As with the non-rotating model, we follow the
collapse until the material is compressed down to the radius of the
marginally stable circular orbit (e.g., Shapiro \& Teukolsky 1983):
\begin{equation}
r_{\rm ms} = M_{\rm BH} \left\{3+Z_2- \sqrt{ \left ( 3-Z_1 \right )
 \left (3+Z_1+2Z_2 \right) } \right\}
\end{equation}
where
\begin{equation}
Z_1 \equiv 1+\sqrt[3]{1-\frac{a_{\rm BH}^2}{M_{\rm BH}^2}}
\left\{\sqrt[3]{1+\frac{a_{\rm BH}}{M_{\rm BH}}}+
\sqrt[3]{1-\frac{a_{\rm BH}}{M_{\rm BH}}}\right\},
\end{equation}
\begin{equation}
Z_2 \equiv \sqrt{3a_{\rm BH}^2/M_{\rm BH}^2 + Z_1^2},
\end{equation}
and $a_{\rm BH}/M_{\rm BH}$ and $M_{\rm BH}$ are the dimensionless
angular momentum and mass of the black hole.  At any given radius in
the proto-black hole, we calculate both $a_{\rm BH}/M_{\rm BH}$ and
$M_{\rm BH}$ and can thus determine when that radius falls below the
marginally stable orbit.  When this happens, we stop the simulation
and remove all of the particles within that radius.  The total mass of
these particles forms the initial black hole mass.  In this manner,
although we do not physically model the collapse to a black hole, we
get roughly both the correct time of collapse and the correct initial
black hole mass (\Tab{300}).

Or do we?  Recall that our collapsing star has a large amount of
angular momentum.  As it collapses, local conservation of angular
momentum causes it to spin up.  As long as the rotational energy is
less than $\sim14\,\%$ of the gravitational potential energy, the
proto-black hole is stable against triaxial deformation.  However,
when $T/|W| > 0.14$, instabilities can occur (e.g. Shapiro \&
Teukolsky 1983).  By plotting the energy ratios ($T/|W|$) as a
function of radius for the proto-black hole 0.5\,\Sec, and 0\,\Sec
before black hole collapse, we see that much of the rotating
proto-black hole is unstable to triaxial deformations
(\Fig{rotation}).  The growth time for such instabilities is (Schutz
1983):
\begin{equation}\label{eq:growth}
\tau \sim \frac{T/|W|}{\Omega} \left ( \frac{\Omega r}{c} 
\right )^{-5}
\end{equation}
where $\Omega$ is the rotational velocity at radius $r$.
\Figure{rotation} also shows the instability growth time versus mass
0.5\,\Sec and 0\,\Sec before the collapse of the proto-black hole.
According to our analysis, instabilities would have time of grow in
our rotating models, possibly forming smaller clumps which would
collapse and then merge to form the central black hole (Bond et al. 1984).  
By comparing the growth time to the time to collapse for both our 
rotating models, we can estimate the range of masses in the proto-black 
hole which might develop these instabilities (\Tab{300}).  \Equation{growth} is
not sufficiently accurate to decisively prove that such instabilities
will develop before the proto-black hole collapses and such
verification awaits three-dimensional simulations which can model the
star over tens to hundreds of orbits.  Even if these instabilities
occur, they will not affect the accretion onto the black hole
significantly (\Sect{accretion}) because the individual clumps will
merge and form a central black hole at roughly the same time as our
assumed stable model.

\pFig{rotation}
\ifthenelse{\boolean{emul}}{
\vspace{1.5\baselineskip}
\noindent
\includegraphics[width=\columnwidth]{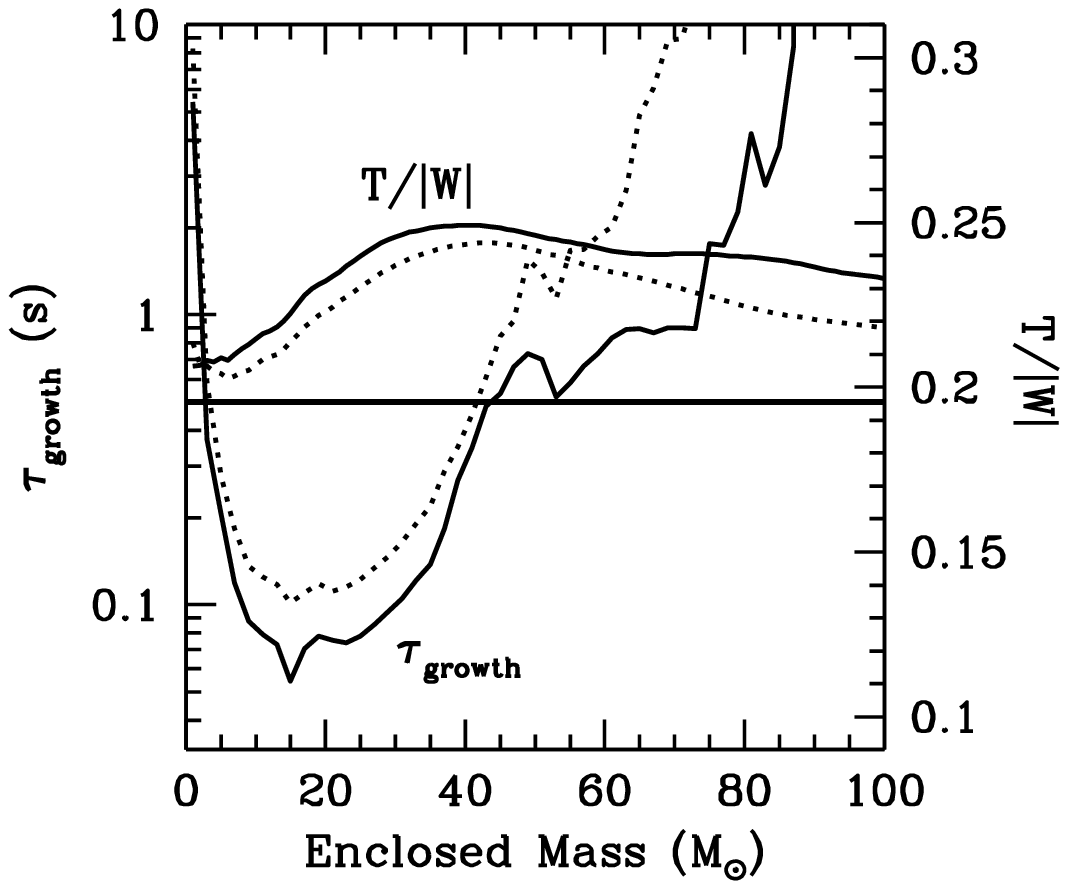}
\figcaption{\Figrotation}
\vspace{1.5\baselineskip}}{}

Although the qualitative picture we have described is the same for
both rotating models (B and C), the actual formation time of the black
hole differs considerably.  This is because in Model C, angular
momentum is transported out of the core using the $\alpha$-disk
prescription.  This lowers the spin of the black hole ($a_{\rm BH}$)
which increases the radius of the marginally stable orbit ($r_{\rm
ms}$) and ultimately causes the black hole to form more quickly.  By
choosing a high disk viscosity ($\alpha_{\rm disk}=0.1$), our Model C
estimates the extreme effect of viscosity and the true answer probably
lies between Models B and C.  Fortunately, these differences do not
affect our quantitative estimates of gravitational wave emission or
GRT  energies (\Sects{gwaves} and \Sectff{accretion}).

\section{Gravitational Waves}
\lSect{gwaves}

Even if the core does not break into several pieces, because $T/|W|$
exceeds the secular instability criterion ($\sim 0.14$), the star may
still develop a bar-like configuration.  Gravitational wave emission
for rotating bodies in the context of supernova collapse has been
studied in great detail (see Zwerger \& M\"uller 1997 or Rampp,
M\"uller, \& Ruffert 1998 for reviews).  We will use here the
expressions derived by Zwerger \& M\"uller for the quadrupole wave
amplitude, the gravitational wave field ($h_+$) and the total
gravitational energy emitted (equations 20-22 of Zwerger \& M\"uller
1997).  In all of our simulations, $|A^{\rm E2}_{20}|$ peaks near
1.5\E7\,\cm, nearly 4 orders of magnitude greater than most of the
simulations of core-collapse supernovae by Zwerger \& M\"uller (1997).
The amplitude of the corresponding waveform is also high:
$h_+=1.3\E{-21}/d(Gpc)$.  Recall that 300\,\Msun stars are only likely
to form at very high redshifts ($z \approx 15$), corresponding to
distances beyond 7.5\,\Gpc for $H_0 \approx
60\,\km\,\Sec^{-1}\,\Mpc^{-1}$. For most of these stars, $h_+ \approx
\Ep{-22}$.  The total energy emitted in gravitational waves during the
first few seconds of collapse is $\sim \Ep{-3}\,\Msun\,c^2$.  During this
time, a 100\,\Msun black hole forms and roughly \Ep{-3}\,\% of its
rest mass energy is converted into gravitational waves.  Like the 
neutrino emission, our calculation of the energy lost to gravitational 
waves is much smaller than that predicted by Bond et al. (1984).  
However, if the core of the proto-black hole breaks up into pieces, the 
gravitational wave emission could be much larger.

We have only scratched the surface of our understanding of the
collapse of these massive stars, but already they show some potential
as sources of powerful gravitational waves.  Unfortunately, the
development of secular instabilities requires the ability to model the
proto-black hole over many orbital periods in three dimensions.  While
we can follow the collapse all the way to black hole formation in two
dimensions, the development of instabilities (and a more accurate
estimate of gravitational wave emission) awaits future
three-dimensional simulations. We would be happy to provide any of our
models to researchers who would like to attempt this problem.

\section{Black Hole Accretion and Gamma-Ray Transients}
\lSect{accretion}

Once a black hole has formed, the rest of the dense inner core
quickly collapses inside, but in the rotating models, some material
along the equator is slowed by centrifugal force.  If there is
sufficient angular momentum to form an accretion disk, this material
will be slowed and pile up outside the hole. The disk that forms would
be similar to those studied by Popham et al. (1999).  Such black hole
accretion disk systems are believed to power gamma-ray bursts, either
through neutrino annihilation or magnetic field powered jets.  To
assess whether the collapse of 300\,\Msun stars might produce some
sort of gamma-ray transient, we must first calculate the disk and
black hole properties.

\pFig{disk}
\ifthenelse{\boolean{emul}}{
\vspace{1.5\baselineskip}
\noindent
\includegraphics[angle=90,width=\columnwidth]{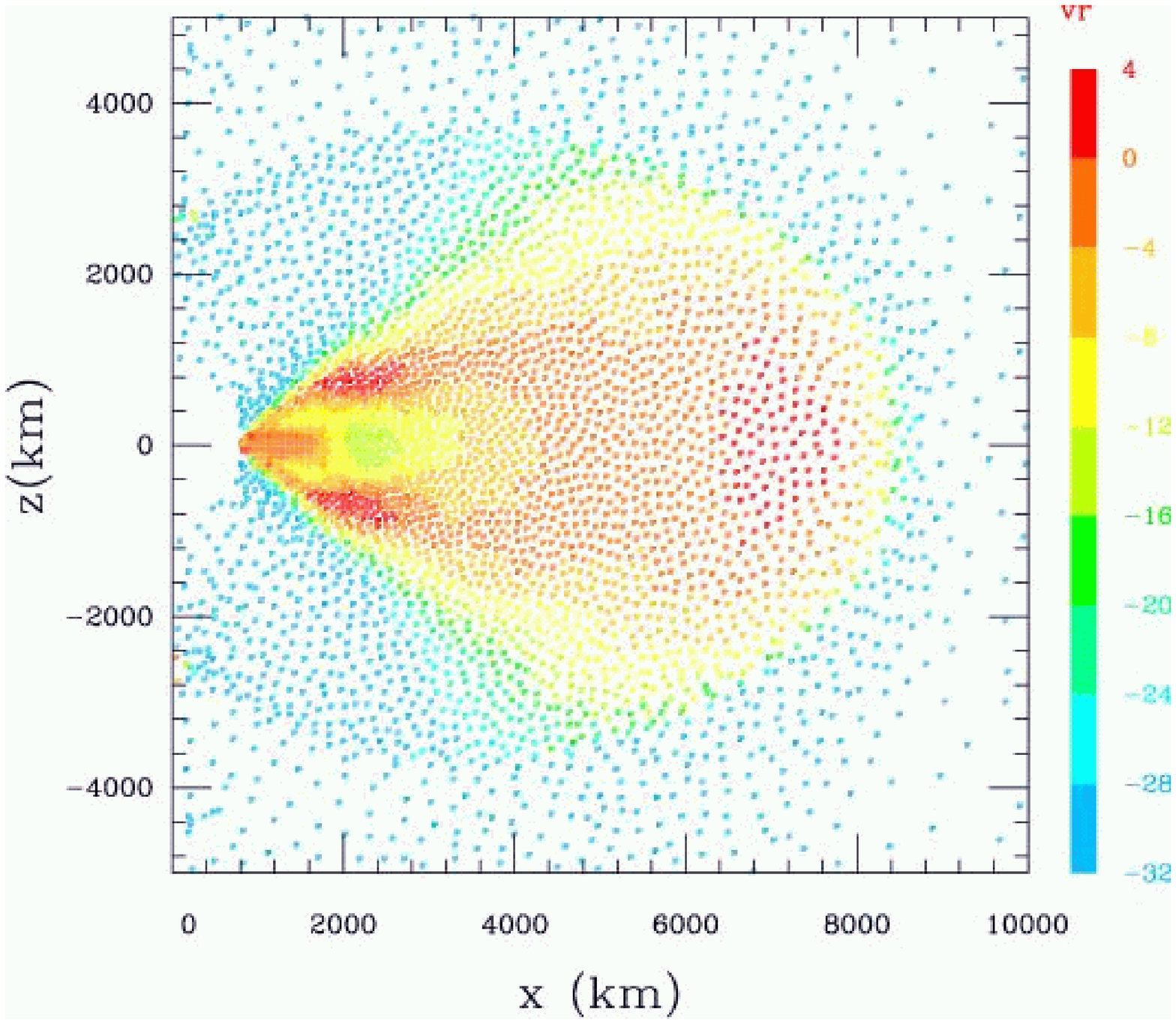}
\figcaption{\Figdisk}
\vspace{1.5\baselineskip}}{}

We follow the accretion into the black hole by placing an absorptive
inner boundary in the core at the marginally stable orbit.  As
particles move inside the marginally stable orbit ($r_{\rm ms}$),
their mass and angular momentum is added to the black hole.  The
boundary condition corresponding to the last stable orbit also expands
as the black hole gains mass.  For the non-rotating model, almost all
of the potential energy of the accreting material was carried into the
black hole (recall \Fig{NeutNR}).  Even in the rotating models, much
of the star collapses onto the black hole before a stable accretion
disk can form.  When angular momentum is conserved locally (Model B,
\Fig{disk}), a stable disk forms only after roughly 140\,\Msun of the
star has accreted (\Fig{disk}).

To produce a GRT, energy generated in the disk and transported, either
by neutrino emission or magnetohydrodynamical processes to matter
above and below the black hole along the rotational axis, must generate
sufficient pressure and momentum to reverse the implosion of material
accreting along the poles (MacFadyen \& Woosley, 1999). The evolution
of the polar accretion and disk accretion rates and the efficiency for
this transport thus dictate if and when an explosion is likely to
occur.  \Figure{accretion} shows the total accretion rates for our three
simulations along with an analytic estimate assuming the mass falls in
at the free-fall time:
\begin{equation}
t_{\rm ff}=\frac{\pi}{2} \sqrt{\frac{ r_0^3}{G M_{\rm enclosed}}}.
\end{equation}
This analytical estimate does not take into account pressure forces or
angular momentum, both of which affect the accretion rate.  The
accretion rate in the non-rotating model shows a large dip in the
accretion rate over the mass range between $\sim40-80\,\Msun$.  This
occurs because nuclear burning injects energy into the collapsing
material and slows its infall.  This effect is not seen in the
rotating models because the material in the proto-black hole burns
into iron elements before the formation of the black hole.  Later,
after black hole formation, the material does not have this additional
reservoir of energy.  The accretion rate decreases at lower black hole
masses in the rotating models because some of the material has enough
angular momentum to support itself in a disk.

\pFig{accretion}
\ifthenelse{\boolean{emul}}{
\vspace{1.5\baselineskip}
\noindent
\includegraphics[width=\columnwidth]{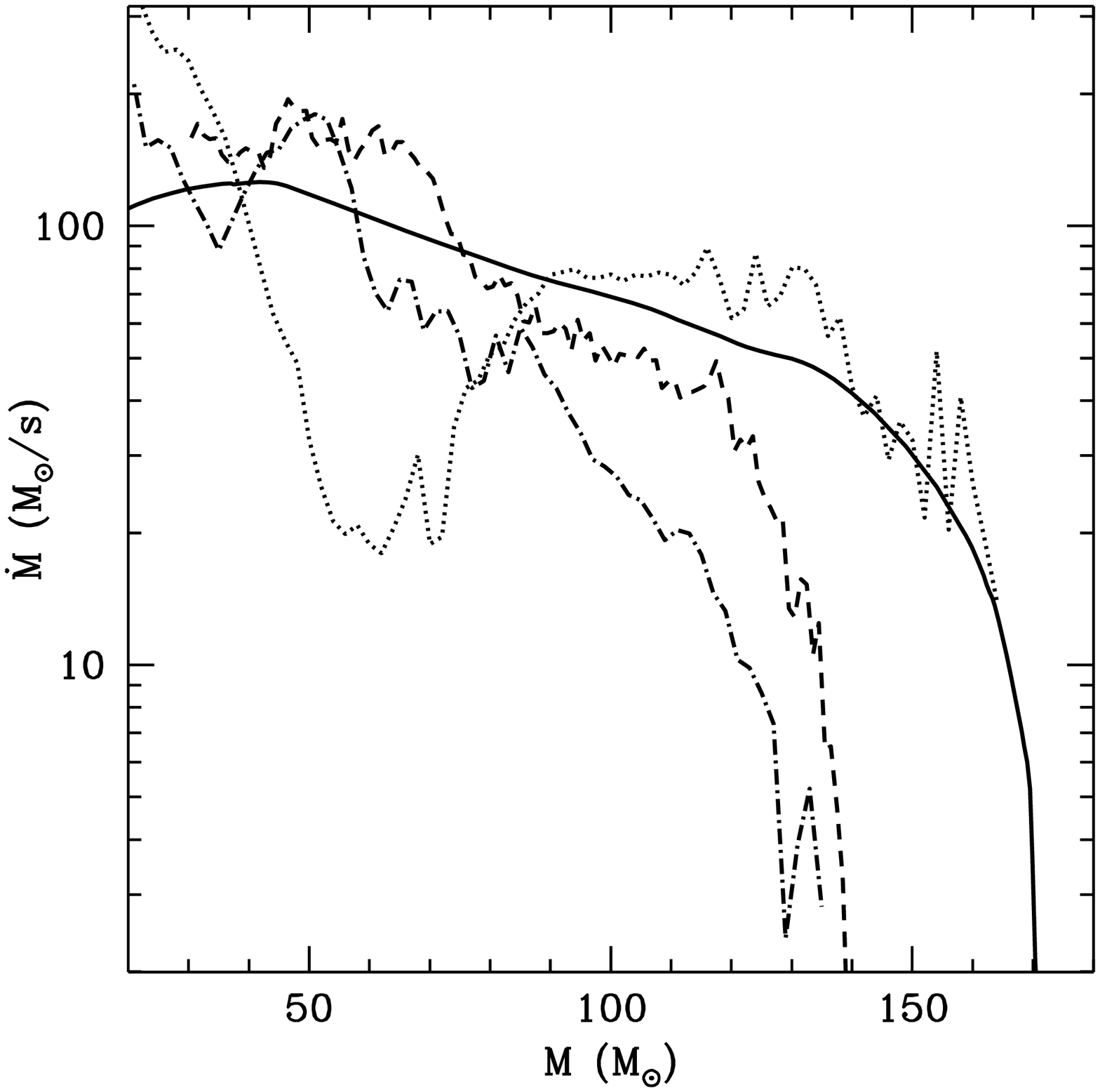}
\figcaption{\Figaccretion}
\vspace{1.5\baselineskip}}{}

In both of our rotating simulations, a disk forms in a wedge extending
up to 35-40$^\circ$ above the equator.  When we include angular
momentum transport (Model C), the angular momentum in the outer layers
of the helium core increases (as these layers gain angular momentum 
from the core), and the disk in Model C is initially
larger both in extent and mass than in the conserved case (Model B).
Although the material in the disk falls much slower into the black
hole than along the poles, the density in the disk quickly becomes so
much higher than in the polar regions that it dominates the accretion
onto the black hole (beyond a black hole mass of 120\,\Msun, disk
accretion makes up $>70\,\%$ of the black hole accretion).  By the
time the total accretion rate drops below 10\,\Msuns, the accretion
along the polar region (the entire region beyond the 40$^\circ$ disk)
accretes less than 3\,\Msuns and this rate is dropping rapidly.
Unfortunately, our limited resolution makes it difficult for us to
follow the accretion rate when it falls below a few \Msuns.  In Model
B (where the angular momentum is locally conserved), the accretion
rate in the disk will drop as abruptly as the rate in the poles as the
disk becomes entirely supported by centrifugal forces.  However, when
we include angular momentum transport (Model C), the disk will
continue to accrete at a rate of $\sim 1-10\,\Msuns$ (for an
$\alpha_{\rm disk}=0.1$) until the entire $\sim40\,\Msun$ disk has
accreted onto the black hole.

It is unlikely that any jet will form before the polar regions clear
and the accretion rate along the poles drops below some critical rate.
This critical rate depends upon the mechanism used to power the
explosion.  For the neutrino-driven mechanism, the accretion rate in
the polar region must drop below 1-10\,\Msuns so that the infalling
material is optically thin to neutrinos.  When the total accretion
rate falls below 10\,\Msuns, corresponding to an accretion rate in the
50$^\circ$ polar region of 2-3\,\Msuns, the conditions around the
black hole may drive a GRB explosion.  At this time, we can compare
the properties of our black hole accretion disk systems with those
models produced by Popham et al. (1999).  The black hole mass and spin
for both our rotating models are listed in \Tab{300}.  The total disk
mass available for rapid accretion is roughly $170\,\Msun-M^f_{\rm
BH}$.

Popham et al. (1999) considered black hole masses between 3 and
10\,\Msun.  The efficiency at which neutrino annihilation converts the
gravitational potential energy released into fireball energy drops
dramatically as the black hole mass increases.  As the event horizon
increases, a) the density and temperature in the disk decreases,
decreasing the total neutrino flux and neutrino energy and b) the
volume of the annihilation region increases, reducing the neutrino
density.  From Popham et al. (1999) we find that the neutrino
annihilation conversion efficiency goes roughly as $M_{\rm BH}^{-n}$
where $n$ is roughly 2-3.  Even if the disk accretes at a rate of
10\,\Msuns, the energy deposited by neutrino annihilation
will not be high enough to drive an explosion.  Note that in our
simulations, the accretion rate through the disk in all of our
simulations is less than 10\,\Msuns when the polar region has
cleared.

Jets driven by magnetic fields may produce a strong explosion though.
It is difficult to make any quantitative predictions about
magnetic-field driven GRBs, simply because the exact mechanism is not
well understood.  However, by the time the total accretion rate has
dropped below 10\,\Msuns, a well-defined disk has formed, which is
required for most of the GRB engines driven by magnetic fields.  The
total potential energy available to drive an explosion is
$\epsilon_{\rm spin} \, \epsilon_{\rm MHD} \, M_{\rm disk} \, c^2$,
where the potential efficiency of the accretion disk ($\epsilon_{\rm
spin}$) is $\sim 0.11$ for $a/M=0.74$.  Assuming that the efficiency
of our magnetic field driven explosion ($\epsilon_{\rm MHD}$) is 0.1,
the yield from Models B and C are 6\E{53} and \Ep{54}\,\erg,
respectively.  Beamed into 1\,\% of the sky, these bursts would have
inferred isotropic explosion energies of nearly \Ep{56}\,\erg, easily
visible at redshift 10 and roughly an order of magnitude more
energetic than ordinary collapsars.

The time scale for these bursts can be crudely estimated from the 
$\alpha$-disk model (Shakura \& Sunyaev 1973), $\tau_{\rm visc} \sim
r^2/(\alpha H^2 \Omega_K)$, with $\alpha$, the disk viscosity
parameter, $\Omega_K$, the Keplerian angular velocity, and $r$ and $H$
the radius and thickness of the disk, respectively. For $r \sim 3 H
\sim 5000\,\km$ (\Fig{disk}) and $\alpha \sim 0.1$, the time scale is
very roughly 10\,\Sec, but there is considerable uncertainty in both
$\alpha$ and the other parameters of this equation. That this value is
approximately equal to the duration of many common GRBs is probably
coincidental, especially given the large redshift and time dilation we
expect for 300\,\Msun stars at death (\Sect{preSN}).  However, it does
suggest the possibility of other forms of GRTs with time scales of
perhaps minutes and a hard x-ray spectrum. Such an energetic jet would
also likely disrupt the star in a gigantic ``hypernova'' explosion.

If it does not, then the outer part of the hydrogen envelope will
accrete in about a day and has enough angular momentum to form a disk
(\Fig{300}). This may make an even longer fainter transient of some sort 
and may be an important source of X-ray photons in the early universe.

\acknowledgements
This research has been supported by NASA (NAG5-2843,
MIT SC A292701, and NAG5-8128), the NSF (AST-97-31569), the US DOE
ASCI Program (W-7405-ENG-48), and the Alexander von Humboldt-Stiftung
(FLF-1065004).  It is a pleasure to thank Markus Rampp for advice on 
calculating gravitational wave emission.

{}

\begin{deluxetable}{lcccccccc}
\tablewidth{40pc}
\tablecaption{Collapse of 300\,\Msun Star}
\tablehead{ \colhead{Model\tablenotemark{a}} & \colhead{$M_{\rm Unstable}$}  
& \colhead{$T_{\rm Coll}$\tablenotemark{b}} & 
\colhead{$M^i_{\rm BH}$} & \colhead{$a^i_{\rm BH}/M^i_{\rm BH}$} & 
\colhead{$M_{\rm Proto-BH}$} & \colhead{$t^{\rm disk}$}
& \colhead{$M^{\rm disk}_{\rm BH}$} & 
\colhead{$a^{\rm disk}_{\rm BH}/M^i_{\rm BH}$} \\
\colhead{} & \colhead{(\Msun)} & \colhead{(s)}  & \colhead{(\Msun)} 
& \colhead{} & \colhead{(\Msun)} & \colhead{(s)\tablenotemark{c}} 
& \colhead{(\Msun)\tablenotemark{c}} & \colhead{\tablenotemark{c}}}

\startdata

A & - & 1.75 & 13 & 0 & - & - & - & - \\
B & 5-40 & 3.7 & 35 & 0.73 & 90 & 3.5 & 136 & 0.74 \\
C & 12-35 & 2.8 & 13 & 0.55 & 70 & 3.6 & 122 & 0.72 \\

\tablenotetext{a}{Model A neglects rotation.  Model B assumes the
initial rotation of the progenitor model and conserves locally the
angular momentum for the duration of the simulation.  Model C uses the
same initial rotation as Model B, but evolves the angular momentum
using an $\alpha$-disk prescription ($\alpha_{\rm d}=0.1$).}
\tablenotetext{b}{Time elapsed after the simulation is mapped from the
KEPLER output.}  \tablenotetext{c}{The ``disk'' quantities (last 3
columns) refer to the time, black hole mass, and black hole rotation
at which the accretion rate drops below 10\,\Msuns.}

\enddata
\lTab{300}
\end{deluxetable}
\clearpage

\ifthenelse{\boolean{emul}}{}{

\clearpage
\onecolumn

\ifthenelse{\boolean{\IncludeFigures}}{
\renewcommand{\figcaption}[2][]{
\clearpage
\begin{figure}
\epsscale{0.8}
\plotone{#1}
\caption{#2}
\newpage
\end{figure}
}}{}

\figcaption[\FigTwoHundredFiftyFile]{\FigTwoHundredFifty}
\figcaption[\FigThreeHundredFile]{\FigThreeHundred}
\figcaption[\FigDensNRFile]{\FigDensNR}
\figcaption[\FigNeutNRFile]{\FigNeutNR}
\figcaption[\FigforceFile]{\Figforce}
\figcaption[\FigvelFile]{\Figvel}
\figcaption[\FigpBHFile]{\FigpBH}
\figcaption[\FigdensFile]{\Figdens}
\figcaption[\FigneutrinoFile]{\Figneutrino}
\figcaption[\FigrotationFile]{\Figrotation}
\figcaption[\FigdiskFile]{\Figdisk}
\figcaption[\FigaccretionFile]{\Figaccretion}
}

\end{document}